\documentstyle[11pt,paspws98,psfig]{article}
\def\footnoterule{\hrule}

\newcommand{\citeauthor}[1]{\def\citeauthoryear##1##2##3{\rm ##1}\cite{#1}}
\newcommand{\citeyear}[1]{\def\citeauthoryear##1##2##3{\rm ##3}\cite{#1}}
\newcommand{\citeN}[1]{\citeauthor{#1} (\citeyear{#1})}
\newcommand{\citeNP}[1]{\citeauthor{#1} \citeyear{#1}}

\begin{document}

\title{Origin and Propagation of Fluctuations of Turbulent Magnetic Fields} 
\footnote{\small 
In: High Resolution Solar Physics: Theory, Observations, and Techniques
 (Rimmele T., Balasubramaniam K. S. \& Radick R. eds.), Proc. 19th NSO/SP
 Workshop, {\it Conf. Series Astron. Soc. Pacif.}, in press}
\author{K. Petrovay}
\affil{\it Instituto de Astrof{\'\i}sica de Canarias, 
 La Laguna (Tenerife), E-38200 Spain}

\markboth{\sc K. Petrovay}
         {\sc Fluctuations of Turbulent Magnetic Field}

\begin{abstract}
The degree of linear polarization has recently been found to show wide random
variations over the solar disk. These variations are presumably at least partly
due to fluctuations in the flux density of turbulent photospheric
magnetic fields and associated variations in the degree of Hanle 
depolarization. In order to understand the origin of such large scale
fluctuations of the turbulent magnetic flux density we develop a 
phenomenological model 
to calculate the spatial Fourier spectrum of the fluctuations of turbulent
magnetic fields in the solar photosphere and convective zone. It is found that
if the model parameters are fitted to turbulence closure models and numerical
experiments the characteristic scale of the fluctuations is by about an order
of magnitude larger than the turbulence scale (the scale of the granulation),
owing to the more effective quenching of small-scale fluctuations by turbulent
diffusion. 
\end{abstract}


\section{Introduction}
In a highly conductive and strongly turbulent plasma like the solar photosphere
random small-scale deviations of the flow field from mirror symmetry are
expected to incessantly generate small magnetic flux loops of random 
orientation. This is a miniature analogue of the large-scale dynamo process 
occurring in a globally non-mirror-symmetric flow, for which reason it is known
as small-scale dynamo action (\citeNP{Leorat+:smalldyn.closure}, \citeNP{Kida+}, 
\citeNP{Petrovay+Szakaly:AA1}). Owing to the random orientation of the
loops, no net large-scale field will arise, but a non-zero mean magnetic energy
density $B^2$ and mean unsigned flux density $|B|$ results. Most of this
turbulent flux resides in magnetic structures with scales much smaller than 
the characteristic scale of the turbulent velocity field (which is the granular
scale of $\sim 1000\,$km in the solar case). Traditional Zeeman
magnetography is ``blind'' to these fields as the net circular polarization
of the mixed polarity small-scale field cancels out in a resolution element.

It has long been recognized that the Hanle effect (magnetic depolarization of
linearly polarized radiation) may offer a way to detect the turbulent fields
(\citeNP{Stenflo:Hanlelimit}). The observational study of turbulent fields has
however been hampered by the shortage of observational data, by the lack of a
reliable radiative transfer theory for polarized radiation in magnetic fields 
and by the fact that
beside turbulent fields, the Hanle effect is also due to resolved magnetic
elements (network, ephemeral active regions) and to the overlying canopy fields
(especially for lines formed higher in the atmosphere). Nevertheless, in recent
years important advances have been made in all these areas (\citeNP{Faurob},
\citeNP{Faurob+me}, \citeNP{Bianda+:Ca}, \citeNP{Bianda+:Sr}, 
\citeNP{Landi:dens.mx}, \citeNP{JTB+Landi}, \citeNP{Landi:Nature}).

On of the most intriguing recent discoveries is the finding of
\citeN{Stenflo+:Hanle.fluct} that the degree of linear polarization $Q/I$ shows
large amplitude random variations over the solar disk. While, as mentioned
above, canopy fields, network elements and ephemeral active regions may also
contribute to the observed Hanle depolarization, it is still likely that at
least part of this observed spatial variation is due to the presence of 
similar fluctuations in the flux density $|B|$ of the turbulent photospheric
magnetic field.

The presence of fluctuations should not come as a surprise from a theoretical
point of view. After all, in the turbulent solar photosphere any physical
quantity should show fluctuations, and even variations of an amplitude
comparable to the mean value are commonplace. What is more surprising is the
\it spatial scale \/ \rm of the fluctuations. While the observations have a
resolution of about $1"$ along the slit, the dominant variations seem to 
occur on the much
larger scale of $\sim 10^4\,$km. Variations on the granular scale are of much
smaller amplitude. (See e.g.\ Fig.\ 3 in \citeNP{Stenflo+:Hanle.fluct}.) The
real theoretical challenge is therefore to understand how the dominant scale of
fluctations of turbulent flux density can be so much larger than the
turbulence scale? 

One popular explanation for the existence of large-scale photospheric 
structures (such as supergranulation) is that they are the ``imprints'' of
processes going on deeper down in the convective zone where the characteristic
scales (determined by the pressure scale height $H$) are larger. Alternatively,
it is of course also possible that the large scales are due to some local
photospheric process like an inverse cascade. In order to resolve this problem,
we need a model for the generation and transport of turbulent magnetic flux
that takes into account both the generation and saturation processes constituting
the small-scale dynamo and the turbulent transport of $|B|$ throughout the 
underlying convective zone. In the following such a model will be presented.

\section{The spectrum of fluctuations: a linear model}
The evolution equation for the unsigned flux density $|B|$ of the turbulent
field should read something like 
\begin{equation}
\partial_t|B|=
\underbrace{\nabla(\beta\,\nabla|B|)}_{\mbox{\scriptsize turbulent diffusion}}
+\underbrace{|B|/\tau_+}_{{\begin{array}{c}\mbox{\scriptsize linear generation} \\
\mbox{\scriptsize (small-scale dynamo)}\end{array}}}
-\underbrace{(|B|/B_0)^q\,|B|/\tau_-}_{\begin{array}{c}\mbox{\scriptsize nonlinear 
 saturation} \\ \mbox{\scriptsize $q>0$}\end{array}}
\end{equation}
As the transport of the highly intermittent magnetic field in a turbulent
plasma is due to advection of flux tubes irrespective of their polarity, the
transport terms are expected to be identical to those for the signed field 
$B$. For simplicity, here we consider an isotropic turbulent diffusion with
diffusivity $\beta=lv/3$ (first term on the r.h.s.), $l$, and $v$ being the
correlation length and r.m.s. amplitude of the turbulent velocity field. The
linear generation term corresponding to a small scale dynamo would lead to an
exponential growth of a homogeneous $|B|$ field to infinity, were it not for a
higher order term leading to the saturation of $|B|$ at a finite value $B_0$,
induced by the curvature force (last term on r.h.s.). 
Numerical simulations and closure calculations (\citeNP{Kida+},
\citeNP{Nordlund+:undershoot}, \citeNP{Durney+:basal}, \citeNP{DeYoung}) show
that $B_0$ is about an order of magnitude lower than the equipartition flux
density: $B_0\sim B_{\mbox{\scriptsize eq}}/10$. 

As the coefficients $1/\tau_+$ and $1/\tau_-$ are functionals of the turbulent
velocity field they are expected to show considerable fluctuations around their
mean values. Hence, the flux density $|B|$ determined by the equilibrium of
generation and nonlinear saturation processes will also fluctuate around its
mean value $\overline{|B|}\simeq B_0$. It is plausible to assume
\begin{equation}
\overline{1/\tau_-}=\overline{1/\tau_+}=1/\tau\equiv v/l . 
\end{equation}
Furthermore, introducing the notation $|B|/B_0=1+b$, it greatly simplifies
the treatment if one assumes $b\ll 1$. (In Section 3 we will consider the
problem to what extent this simplification affects the results.)
This allows us to linearize equation (1). In the case when $b$ varies much 
faster with depth than $B_0$ we obtain
\begin{equation}
\partial_tb={\nabla(\beta\nabla b)}-qb/\tau+(1/\tau_+-1/\tau_-)'
\end{equation}
Now the second term on the r.h.s.\ describes the net mean restoring effect of
generation and saturation terms, tending to reduce the fluctuation $b$, while 
the last term is the \it fluctuation generation term\/ \rm or \it forcing term
\rm arising owing to the turbulent fluctuations in the coefficients. 

We further introduce the
(physically plausible) assumption that the Fourier spectrum of this forcing 
term is dominated by those isotropic modes whose vertical phase is
such that at the depth where the pressure scale height equals the inverse of 
their horizontal wavenumber $k$ they are maximal:
\begin{equation} (1/\tau_+-1/\tau_-)'= \frac1\tau\sum_{k}{
   \hat b_f(k,z)\,\exp[i(kx+ky+\omega t)]}
\end{equation}
(Note that throughout this paper we take the logarithmic pressure $\ln P$ as
independent variable in the vertical direction, the depth $z$ being just a
shorthand notation for a function $z(\ln P)$, determined by a convection zone
model. $x$ and $y$ are the horizontal coordinates.) 
For the solution of equation (3) we take a similar {\it Ansatz}:
\begin{equation} b= \sum_{k}{ 
   \hat b(k,z)\,\exp[i(kx+ky+\omega t)]}
\end{equation}

Let us consider one mode only. (Note that in this case $\hat b$ can always be
considered real as by virtue of our assumption about the vertical phase of
modes $\hat b(k,z)$ may be written as $\hat b_1(k)\,\cos[\ln P-\ln P_0(k)]$,
and  the initial phase in $\hat b_1$ can be chosen freely by a displacement of
the time axis.) We substitute (4) and (5) into (3), simplify, and take the 
real part. To simplify the notation, from this point onwards we omit the hats.
With this we arrive at
\begin{equation}
   -d_z(\beta\,d_z b)+(2k^2\beta+q/\tau)\,b=b_f/\tau
\end{equation}
This equation determines the Fourier amplitude $b(z,k)$ of each mode of 
horizontal 
wavenumber $k$ in the spectrum of turbulent magnetic field fluctuations.
Vertical diffusion is now separated in the first term; horizontal diffusion and
the restoring force constitute the second and third terms on the l.h.s. 

The r.h.s.\ is the Fourier amplitude of the forcing. $b_f$ is the fluctuation
amplitude produced in time $\tau$ by the forcing if other terms were not
present. As the forcing is due to the action of the fluctuating flow field on
the existing magnetic field, it is plausible to assume $b_f/\tau \propto 
B_k/B_0\tau_k$, where $B_k$ is the corresponding Fourier amplitude in the
spectrum of $|B|$, and $\tau_k(k,z)$ is the eddy turnover time. This spectrum may be
determined from closure calculations, numerical simulations and observations.
\begin{figure}[hbt]
\centerline{\psfig{figure=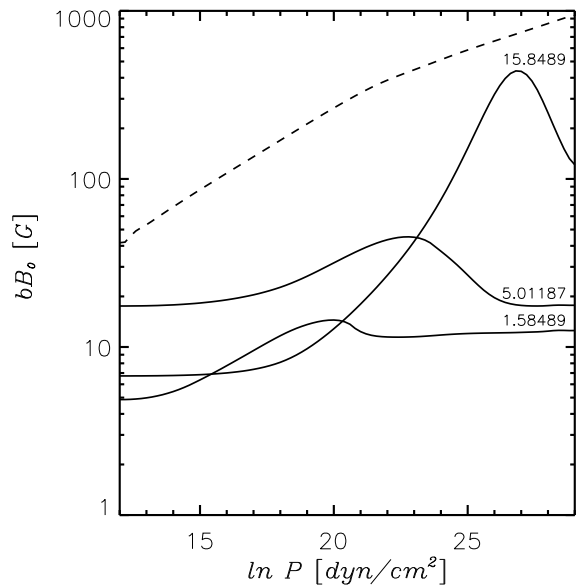,height=8.5 cm}}
\caption{Vertical profiles of some Fourier modes in the spectrum of
fluctuations of the turbulent magnetic field. The modes are labelled by their
value of $1/k$ in Megameter units. Dashed: $B_0$}
\end{figure}
Its simplest representation is by two power laws  joining in a peak at $k\sim
10/l$. Thus, we represent the r.h.s. by
\begin{equation}
b_f(z,k)/\tau(z,k)=10^{2/3-p}\,(k/k0)^p, \qquad k_0\sim 1/H(z), 
\end{equation}
where $p=p_1$ for $k<10k_0$ and $p_2$ otherwise. (Note that in fact a spectrum
with two breaks might be more realistic as the spectrum of $\tau$ is peaked at
a lower wavenumber of $k\sim 1/l$; our purpose here, however, is just to
present a simple example calculation.) On the basis of the observed properties
of photospheric magnetic fields and motion, $p_1\sim 1.5$ seems to be a
realistic choice. The value of $p_2$ will be found to be irrelevant to the
solution below.

For the solution of equation (6) the parameters $\beta$ and $\tau$ are
interpolated from a convective zone model. $q$ is evaluated by fitting a
solution of equation (1), with the diffusive term and the perturbations of the
coefficients neglected, to profiles of
$|B|^2(t)$ resulting from the closure model of \citeN{DeYoung}; this yields 
$q\simeq 0.1$. The boundary conditions may be chosen as closed 
($F\equiv\beta\, d_zb=0$) or open ($F=vb$); experimenting has shown that this
does no exert a very strong influence on the resulting mode profiles. 
For the numerical solution of equation (6) we take $\ln P$ as independent 
variable and use a relaxational method. 

The vertical profiles of some Fourier modes in the spectrum of fluctuations of
the turbulent magnetic field are shown in Figure 1. As expected, the modes with
larger horizontal scales have their maxima in deeper layers. 

\begin{figure}[htb]
\centerline{\psfig{figure=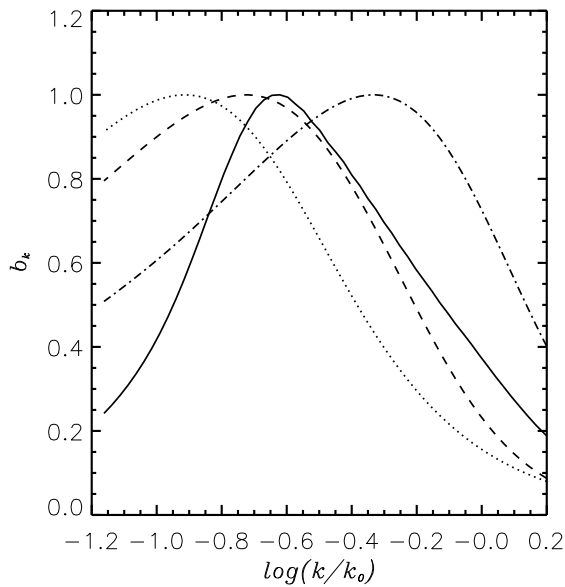,height=8.5 cm}}
\caption{Fourier spectra of the fluctuations of the turbulent magnetic flux
density as functions of the horizontal wavenumber. Solid: $p_1=1.5$, $q=0.1$;
dashed: $p_1=0.5$, $q=0.1$; dash-dotted: $p_1=1.5$, $q=1.0$; dotted: 
$p_1=0.5$, $q=0.1$ with vertical transport switched off}
\end{figure}

Computing a large number of modes and plotting their amplitudes near the
surface against $k$ yields the Fourier spectrum of the
fluctuations in the turbulent magnetic field (Fig.\ 2). Experimenting with
the parameters in equation (6) we find that the choice of $p_2$ is
irrelevant for the spectrum, $B_0/B_{\mbox{\scriptsize eq}}$ 
determines its amplitude, while $p_1$ and $q$ determine the shape of the 
spectrum. A striking feature of the spectra in Figure 2 is that their maxima
fall to significantly lower wavenumbers than $k_0=1/l$, i.e.\ the
characteristic scale of the fluctuations in turbulent magnetic flux density is
much larger than the typical scale of turbulent motions. The physical
background of this phenomenon is that the high wavenumber components are more
efficiently suppressed by diffusion ($2k^2\beta\,b$ term in eq.\ (6)). The
wavenumber of the maximum increases with $q$ (dashed vs.\ dash-dotted curves), 
as a stronger nonlinear
saturation reduces the role of the diffusive terms in the equation. On the
other hand, the spectral amplitude
at even lower wavenumbers depends strongly on the spectral index $p_1$
of the forcing (dashed vs.\ solid curves) ---clearly, the shallower the 
forcing spectrum, the more energy is input at the larger scales.

\section{The effect of nonlinearity}
The results presented in the previous section were computed under the
assumption that the fluctuations of $|B|$ have a small relative amplitude. This
assumption is rather dubious in the light of the large Fourier amplitudes found
in the model (Fig.\ 1). In order to have an idea about the extent to which
nonlinear effects may modify the linear results, in this section we present an
alternative model that calculates the fluctuating field without the assumption
of linearity, at the cost of a strong simplification of the geometry: only
$k=0$ modes are considered. This obviously implies that spatial spectra or
correlation lengths cannot be studied; instead, we will compare the \it
temporal\/ \rm autocorrelation of the fluctuations in the linear and nonlinear
cases. 

\begin{figure}[htb]
\centerline{\psfig{figure=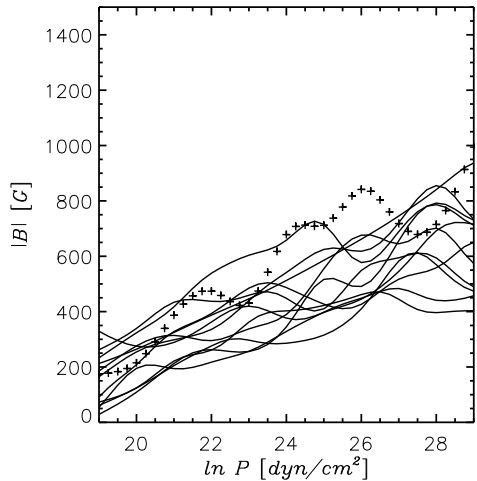,height=7 cm}}
\caption{Snapshots of the solution of equation (7) with $q=1$
at different instants of time. Crosses: initial profile.}
\end{figure}

For $k=0$ modes $d_x=d_y=0$ so we write equation (1) in the form
\begin{equation}
\partial_t|B|=
d_z(\beta\,d_z|B|)+|B|/\tau\left[1-(|B|/B_0)^q\right]+B_f/\tau
\end{equation}
where the last term corresponds to fluctuation forcing by to the turbulent
fluctuations of the velocity field. We model this term as a Gaussian 
stationary random process with correlation time $\tau(z)$, vertical 
correlation ``length'' 1 (in $\ln P$ units), and a mean displacement of 
$0.3 B_0$ over $\tau$ .

Equation (8) is then integrated numerically starting from an arbitrary 
perturbed
initial state. An example solution is presented in Figure 3. Figure 4 shows the
autocorrelation of the fluctuating flux density as a function of the time shift
for different cases. It is apparent (dotted vs.\ dashed curves) that the 
correlation time is primarily
determined by the value of the $q$ nonlinearity parameter, lower $q$ values
corresponding to longer correlation times. This result is a close analogue to
the findings of Section 2 with respect to the spatial correlations. On the
other hand, switching off the diffusive term in equation (8) (dash-dotted 
vs.\ dashed curves) has no significant effect, showing that diffusive quenching
of small spatial scales is \it not\/ \rm the mechanism responsable for the
extended correlations here. Replacing equation (8) with its linear equivalent 
(solid vs.\ dashed curves) does not lead to great modifications in the results.
This may reassure us to some extent of the reliability of the findings of
Section 2 above.

\begin{figure}[htb]
\centerline{\psfig{figure=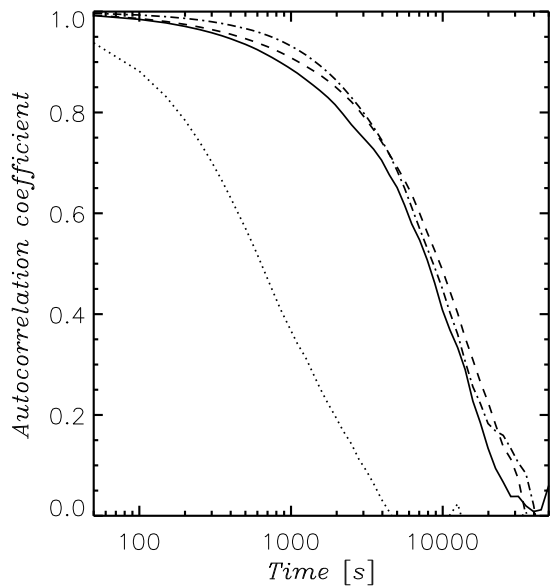,height=8.5 cm}}
\caption{Autocorrelation of the fluctuating flux density as a function of the
time difference at a fixed point in the convective zone in the nonlinear model
with $q=0.1$.
Dashed: reference model; dotted: $q=1$; dash-dotted: reference model with
diffusive term switched off; solid: linear equivalent}
\end{figure}

\section{Conclusion}
The results of Section 2 now enable us to answer the question posed at the end
of the Introduction. Turbulent transport processes indeed result in a dominant
scale for the fluctuations of the turbulent magnetic flux density that is an
order of magnitude larger than the turbulence scale. The physical mechanism
behind this phenomenon is that small-scale fluctuations are more efficiently
damped by diffusion, shifting the spectral peak to lower wavenumbers. The low
value of the $q$ nonlinearity parameter suggested by closure calculations and
simulations also favors larger dominant scales.

On the other hand, the possibility that deeper structures are ``imprinted'' on 
the photospheric pattern can apparently be discarded. Switching off the first
term (vertical diffusion) in equation (6) does not lead to any reduction of the
dominant scales (dotted vs.\ dashed curves in Fig.\ 2; dash-dotted vs.\ dashed
in Fig.\ 4).

All this shows that the observations of strongly varying Hanle depolarization
over the solar disk may be understood from a theoretical point of view even if
all the depolarization is attributed to turbulent fields. Further advances on
both the observational and theoretical side may make more detailed comparisons
between the models and the observations possible, thereby offering the prospect
of a direct observational diagnostics of the properties of the turbulent
dynamo. 

\acknowledgements
This work was funded by the DGES grant no.~95-0028.



\vfill\eject

\end{document}